\documentclass[11pt]{article}
\usepackage{amstex}
\usepackage{oldgerm}

\date{}

\makeatletter

\begin{document}

\title{{\bf Gauge-invariant magnetic perturbations in perfect-fluid
cosmologies}}
\author{Christos G. Tsagas\thanks{Current address: Relativity and Cosmology
Group, School of Computer Science and Mathematics, University of Portsmouth,
Portsmouth PO1 2EG, UK.} 
and John D. Barrow\thanks{e-mail address: j.d.barrow@sussex.ac.uk}\\
{\small {\it Astronomy Centre, University of Sussex,}}\\
{\small {\it Brighton BN1 9QJ, U.K.}}}
\maketitle

\begin{abstract}
We develop further our extension of the Ellis-Bruni covariant and
gauge-invariant formalism to the general relativistic treatment of density
perturbations in the presence of cosmological magnetic fields. We present
detailed analysis of the kinematical and dynamical behaviour of perturbed
magnetized FRW cosmologies containing fluid with non-zero pressure. We study
the magnetohydrodynamical effects on the growth of density irregularities
during the radiation era. Solutions are found for the evolution of density
inhomogeneities on small and large scales in the presence of pressure, and
some new physical effects are identified.\\

\end{abstract}

\section{Introduction}
%%%%%%%%%%%%%%%%%%%%%%
In a recent article, \cite{TB}, we examined the behaviour of cosmological
density perturbations in a universe containing a large-scale primordial
magnetic field, by means of the Ellis and Bruni covariant and gauge-invariant
approach \cite{EB}. Our assumptions were: first, that the conductivity of the
medium is infinite; and second, that the background universe, though permeated
by a coherent magnetic field, remains spatially isotropic to leading
order. The first approximation is a standard simplification of Maxwell's
equations, ignoring any large-scale electric field while preserving the desired
coupling between matter and the magnetic field. The second approximation was
introduced at a later stage of our analysis, to allow the direct comparison
between our results and those from previous Newtonian treatments. Starting from
a general, inhomogeneous and anisotropic, cosmological model we provided the
exact, fully non-linear, evolution formulae for all the basic gauge-invariant
variables. These equations are valid irrespective of the field's strength and
can be linearized about a ``variety'' of smooth background universes. In
\cite{TB}, the main objective was to establish a fully relativistic treatment
of magnetized density perturbations. To achieve this, we focussed upon the
dust era and compared our results to those obtained earlier by the Newtonian
treatments of Ruzmaikina-Ruzmaikin, \cite{RR}, and Wasserman, \cite{W}. In
addition to the desired agreement with the non-relativistic analysis, our
method suggested weak corrections to the evolution of density
disturbances. These are generated by both the isotropic and the anisotropic
pressure that the magnetic field introduces into the cosmological model. Our
results confirmed the relative unimportance of the field for the evolution of
superhorizon-sized density disturbances.

The present article extents our study to the radiation era by including the
isotropic pressure of a perfect fluid in the calculations. During this period
the kinematic evolution of the universe follows more complicated patterns than
in the dust era. The sources of the extra complexities are changes in the fluid
motion due to its non-vanishing pressure, which manifest themselves in a number
of ways. For instance, the acceleration of the fluid depends now on pressure
gradients as well as on the spatial variations of the magnetic field, with the
two of them not necessarily acting in the same sense. So, unlike the
pressure-free case, the field vector and the fluid acceleration are generally
not orthogonal. Moreover, the time derivative of the magnetic field is no
longer a spacelike 3-vector. All these factors leave their trace on the
kinematics of the universe and further complicate its dynamic evolution. It
should be emphasized that we do not consider plasma-physics complexities
induced by dissipative stresses in the energy momentum tensor of the fluid. A
separate study of this magnetohydrodynamical problem, both at linear and
non-linear level, is possible due to the conformal invariance of the stress
tensor and can be found in Subramanian and Barrow \cite{SB1}. These studies
provide a mathematical and physical basis for the evaluation of the
observational effects of cosmological magnetohydrodynamics. The existence of
magnetic fluctuations can leave observable traces on the structure of the
microwave background on small angular scales, which might be detected by future
satellite missions if the magnetic field is strong enough to influence the
formation of large scale structure. Subramanian and Barrow \cite{SB2} have
shown that, for a tangled magnetic field with present strength
$H_0\sim3\times10^{-9}G$, one can expect an RMS microwave background anisotropy
signal of order $5\mu K$ or larger, depending on the angular scale. The
anisotropy in hot or cold spots could be several times larger. The formalism we
have developed can also be used to trace the effects of the damping of magnetic
field fluctuations on the photon and neutrino spectra emerging from the
radiation era. On superhorizon scales the field evolves as though
quasi-homogeneous and will create small expansion anisotropies which will
produce anisotropies in the microwave temperature distribution. The existing
observational data allows us to place upper limits of
$6.8\times10^{-9}(\Omega_0h^2)^{1/2}G$ on the present strength of any {\it
uniform} (i.e. spatially homogeneous) component of the magnetic field
\cite{BFS}, \cite{B}\footnote{$\Omega_0$ is the present value of the density
parameter and $h$ is the Hubble constant in units of $100km/sMpc$.}.

In this report we examine the case of a perturbed {\it spatially flat
Friedmann-Robertson-Walker (FFRW)} magnetized universe filled with a single
barotropic perfect fluid and derive the linear equations that determine its
evolution. We consider the evolution of the basic kinematic and dynamic
quantities and the magnetohydrodynamical effects upon them. We show that, in
regions of subhorizon size, gradients in the energy density of the field,
together with those in the fluid, decelerate the universal expansion and act
as sources of positive spatial curvature. It also appears that the field tends
to smooth out the curvature of the underlying 3-surfaces and, depending on the
spatial curvature, it can act as a source of slightly accelerated expansion.

Given the current interest in structure formation, we provide a set of four
linear first order differential equations that governs the linear evolution of
$\Delta$, the scalar variable that determines the gravitational clumping of
matter and describes the formation of structure directly. During the radiation
epoch the energy density of the fluid and that of the magnetic field fall as
the inverse fourth power of the scale factor. As a result, the Alfv\'{e}n
velocity is time-independent and our system of four differential equations
becomes easier to handle. We provide analytic solutions at both limits of the
wavelength spectrum and compare them to those of a non-magnetic universe. We
show that although the density inhomogeneities retain their basic evolutionary
patterns, the role of the field as an agent opposing their growth is clear. In
particular, large-scale perturbations undergo a power-law evolution, similar
to that of the non-magnetized case, but their growth rate is reduced by an
amount proportional to the field's relative strength. At the opposite end of
the spectrum the density contrast continues to oscillate. Here, the extra
magnetic pressure has simply reduced the oscillation period. The conclusion is
that during the radiation era the magnetic effects are just supplementary to
those induced by the pressure of the relativistic fluid. After the radiation
era ends, the field becomes the sole source of pressure in regions exceeding
the associated at the time Jeans length. We find, in agreement with our earlier
conclusions (see \cite{TB}), that any large-scale magnetic influence ceases
completely by the later stages of the dust era, although at earlier times it
could have forced the density contrast to enter a brief oscillatory phase. The
negative role of the field is also confirmed in the subhorizon regions. On
scales between the Jeans length and the horizon, the magnetic field slows the
power-law growth of the inhomogeneities by an amount depending on its relative
strength.

In \cite{TB} we considered a general FRW universe, allowing for spatially open
and closed unperturbed backgrounds.  However, it is important to recognize that
the gauge invariance of the magnetic field perturbations holds if and only if
the underlying spatial sections are flat.  Accordingly, all equations in
\cite{TB} must be linearized about a FFRW universe and every variable
representing spatial curvature should be treated as a perturbation.  As a
result, terms in the linearized formulae of \cite{TB} that contain the
background 3-curvature constant are non-linear and can be dropped at first
order.  This does not effect the results presented there but restricts their
validity to almost-FFRW models.  Notice that the gauge invariance of the
magnetic field gradients still holds within a perturbed Bianchi-I universe due
to the latter's spatial flatness. In Appendix A we give a full account of this
question.

\section{Preliminaries}
%%%%%%%%%%%%%%%%%%%%%%%
\subsection{Kinematic Variables}
%%%%%%%%%%%%%%%%%%%%%%%%%%%%%%%%
Following \cite{Eh} and \cite{E2}, we assume that the average motion of
matter in the universe defines a future directed velocity 4-vector, $u_i$,
corresponding to a {\it fundamental observer} ($u_iu^i=-c^2$), and generates a
unique splitting of spacetime into ``time'' and ``space'' (1+3
decomposition). For any tensorial quantity $T$, the directional derivative
$\dot{T}=T_{;i}u^i=u^i\nabla_iT$ denotes differentiation along the fluid-flow
lines. The second order symmetric tensor $h_{ij}=g_{ij}+u_iu_j/c^2$ projects
orthogonal to $u_i$ onto what is known as the observer's {\it instantaneous 3-D
rest space} $\Sigma_{\perp}$\footnote{At every event along the worldline of a
fundamental observer, $\Sigma_{\perp}$ is the normal to $u_i$ 3-D subspace of
the 4-D space tangent to that event.}. We also introduce
$\mbox{}^{(3)}\nabla_i$, the covariant derivative operator orthogonal to $u_i$
($\mbox{}^{(3)}\nabla_ih_{jk}=0$), by totally projecting the corresponding 4-D
operator. This is not, however, a derivative on a hypersurface unless the
fluid flow is irrotational. Nevertheless, we will call the 3-gradient
$\mbox{}^{(3)}\nabla_i$ ``spatial'' for simplicity.

The kinematic variables are established by decomposing the covariant
derivative of $u_i$ into its spatial and temporal parts. In particular, we
have
\begin{equation}
\nabla_ju_i=\sigma_{ij}+\omega_{ij}+\frac{\Theta}{3}h_{ij}-
\frac{1}{c^2}a_iu_j ,  \label{ui;j}
\end{equation}
where $\sigma_{ij}=\mbox{}^{(3)}\nabla_{(j}u_{i)}-\Theta h_{ij}/3$ is the shear
($\sigma_{ij}u^i=\sigma_{ij}u^j=0$, $\sigma_i^{\hspace{1mm}i}=0)$,
$\omega_{ij}=\mbox{}^{(3)}\nabla_{[j}u_{i]}$ is the vorticity
($\omega_{ij}u^i=\omega_{ij}u^j=0$), $\Theta=\nabla_iu^i$ is the volume
expansion, and $a_i=\dot{u}_i=u^j\nabla_ju_i$ is the acceleration
($a_iu^i=0$). The magnitudes of the shear and the vorticity are
$\sigma^2=\sigma_{ij}\sigma^{ij}/2$ and $\omega^2=\omega_{ij}\omega^{ij}/2$
respectively. The expansion scalar, $\Theta$, defines a representative length
scale ($S$) along the fluid flow by means of $\dot{S}/S=\Theta/3$. In a
non-rotating universe (i.e. when $\omega_{ij}=0$), $u_i$ is a hypersurface
orthogonal field and $\Sigma_{\perp}$ becomes a 3-surface, namely the
instantaneous rest space of all the fundamental observers.

\subsection{Spacetime Geometry}
%%%%%%%%%%%%%%%%%%%%%%%%%%%%%%%
The global geometry of the spacetime is determined by the Riemann curvature
tensor, conveniently expressed by the decomposition
\begin{equation}
R_{ijkq}=C_{ijkq}+
\frac{1}{2}\left(g_{ik}R_{jq}+g_{jq}R_{ik}-g_{jk}R_{iq}-g_{iq}R_{jk}\right)-
\frac{1}{6}\left(g_{ik}g_{jq}-g_{iq}g_{jk}\right)R ,  \label{Wt}
\end{equation}
where $C_{ijkq}$ is the Weyl conformal curvature tensor,
$R_{ij}\equiv R_{i\hspace{1mm}jk}^{\hspace{1mm}k}$ is the Ricci tensor and
$R\equiv R^i_{\hspace{1mm}i}$ is the Ricci scalar. Both the Ricci tensor and
the Ricci scalar are determined locally by matter through the Einstein field
equations. Conversely, the Weyl tensor describes long range gravitational
effects, such as those of tidal forces and gravitational waves. By definition
$C_{ijkq}$ satisfies all the symmetries of the Riemann tensor and is also
trace-free. It decomposes into an electric and a magnetic part according to,
\cite{M},
\begin{equation}
C_{ijkq}=\frac{1}{c^2}\left(g_{ijsr}g_{kqpt}
-\eta_{ijsr}\eta_{kqpt}\right)u^su^pE^{rt}-
\frac{1}{c^2}\left(\eta_{ijsr}g_{kqpt}
+g_{ijsr}\eta_{kqpt}\right)u^su^pH^{rt} ,  \label{dWt}
\end{equation}
where
\begin{equation}
g_{ijkq}\equiv g_{ik}g_{jq}-g_{iq}g_{jk} ,  \label{gijkq}
\end{equation}
$\eta_{ijkq}$ is the totally antisymmetric spacetime permutation tensor and
$E_{ij}$, $H_{ij}$ are respectively the electric and the magnetic components
of the Weyl tensor. The latter have nothing to do with actual electric or magnetic
fields but derive their name from the Maxwell-like equations they comply with
\cite{MB}. Also notice that
\begin{equation}
C_{ijkq}=0\hspace{5mm}\Leftrightarrow\hspace{5mm}
\begin{cases}
E_{ij}=0 ,\\
H_{ij}=0 .
\end{cases}  \label{CEH0}
\end{equation}
 
\subsection{The Electromagnetic Field}
%%%%%%%%%%%%%%%%%%%%%%%%%%%%%%%%%%%%%%
The electromagnetic field is represented by the antisymmetric Maxwell tensor 
$F_{ij}$. This splits into an electric and a magnetic 4-vector respectively
defined by, \cite{E2},
\begin{equation}
E_i=F_{ij}u^j ,\hspace{2cm}\mbox{\rm and}\hspace{2cm}
H_i=\frac{1}{2c}\eta_{ijkq}u^jF^{kq} .  \label{EH}
\end{equation}
The above confirm that $E_iu^i=H_iu^i=0$, which in turn mean that both
fields lie on $\Sigma_{\perp}$; $E^2\equiv E_iE^i$ and $H^2\equiv H_iH^i$
respectively denote the magnitudes of the electric and the magnetic fields.

\subsection{The Material Component}
%%%%%%%%%%%%%%%%%%%%%%%%%%%%%%%%%%%%
As in \cite{TB}, we consider a universe filled with a single perfect fluid of
infinite conductivity (i.e.
$\overline{\sigma}=cE_iJ^i/E^2\rightarrow\infty$, with $J_i$ representing
the current density). We can now drop the electric field from Maxwell's
equations, which reduce to\footnote{the merit of the infinite conductivity
assumption is that, based on Ohm's law, it can accommodate a zero electric
field with non-vanishing spatial currents (i.e.
$h_i^{\hspace{1mm}j}J_j\neq0$). The latter condition is essential if one
wishes to preserve the coupling between matter and magnetic field (see
Appendix B in \cite{TB}).}
\begin{eqnarray}
2\omega^iH_i&=&\epsilon c ,  \label{M1}\\
\eta^{ijkq}u_j\left(a_kH_q-c^2\nabla_qH_k\right)&=&
c^2h^i_{\hspace{1mm}j}J^j ,  \label{M2}\\
\nabla^iH_i-\frac{1}{c^2}a^iH_i&=&0 ,  \label{M3}\\
\left(\sigma^i_{\hspace{1mm}j}+\omega^i_{\hspace{1mm}j}
-\frac{2\Theta}{3}h^i_{\hspace{1mm}j}\right)H^j&=&
h^i_{\hspace{1mm}j}\dot{H}^j ,  \label{M4}
\end{eqnarray}
where $\omega_i\equiv\eta_{ijkq}u^j\omega^{kq}/2c$ is the vorticity vector and
$\epsilon\equiv-J_iu^i/c^2$ is the charge density. For our purposes the last
two equations are the important ones. More specifically, (\ref{M3}) provides
the familiar vanishing 3-divergence law for the magnetic field (i.e.
$\mbox{}^{(3)}\nabla_iH^i=0$), whereas (\ref{M4}), when contracted with the
magnetic field vector, gives
$\sigma_{ij}H^iH^j=2\Theta H^2/3+(H^2)\dot{\mbox{}}/2$ and simplifies the
expression for the energy density conservation law, \cite{TB}.

The energy-momentum tensor for a magnetized single perfect fluid of infinite
conductivity has the form, \cite{TB},
\begin{equation}
T_{ij}=\left(\mu+\frac{H^2}{2c^2}\right)u_iu_j+
\left(p+\frac{H^2}{6}\right)h_{ij}+ \Pi_{ij}  \label{emt}
\end{equation}
with the pressure ($p$) and the mass density ($\mu$), the latter including
contributions from the internal thermal energy, related by a suitable
equation of state. The symmetric, traceless and completely spacelike tensor
$\Pi_{ij}=H^2h_{ij}/3-H_iH_j$ describes the anisotropic pressure induced by
the magnetic field.\footnote{In \cite{TB} we represented the anisotropic
magnetic stresses by $M_{ij}$ instead of $\Pi_{ij}$. Other changes relative to
that article are; $a_i$ has replaced $\dot{u}_i$ as the acceleration vector;
the 3-Ricci scalar has changed from ${\cal K}$ into $K$, while now
${\cal K}\equiv S^2K$; the gradient $\mbox{}^{(3)}\nabla_iH^2$ is represented
by ${\cal H}_i$ and not by ${\cal B}_i$ as in \cite{TB}; and the scalar
${\cal B}$ is no longer the Laplacian $\mbox{}^{(3)}\nabla^2H^2$ but equals the
dimensionless ratio $S^2\mbox{}^{(3)}\nabla^2H^2/H^2$.}

\subsection{Inhomogeneity Variables}
%%%%%%%%%%%%%%%%%%%%%%%%%%%%%%%%%%%%
In a FFRW universe all physical quantities are functions of cosmic-time only,
while the shear, the vorticity, the acceleration, all the anisotropic stresses,
the curvature of the spatial sections and the electric and magnetic components
of the Weyl tensor vanish. Within a nearly-FFRW universe, spatial perturbations
in the energy density and the pressure of the fluid, in the expansion and in
the magnetic field are described by four key variables covariantly defined in
\cite{EB} and \cite{TB}. These are: the comoving fractional orthogonal spatial
gradient of the energy density, $D_i\equiv S\mbox{}^{(3)}\nabla_i\mu/\mu$; the
orthogonal spatial gradient of the pressure,
$Y_i\equiv\kappa\mbox{}^{(3)}\nabla_ip$, where $\kappa=8\pi G/c^4$ is the
Einstein gravitational constant; the comoving orthogonal spatial gradient of
the expansion, ${\cal Z}\equiv S\mbox{}^{(3)}\nabla_i\Theta$; and the comoving
orthogonal spatial gradient of the magnetic field,
${\cal M}_{ij}\equiv\kappa S\mbox{}^{(3)}\nabla_jH_i$, with
${\cal M}_i^{\hspace{1mm}i}=0$, as (\ref{M3}) requires. Each one of these
3-gradients vanishes in a perfect FFRW model (see also Appendix A), thus
satisfying the criterion for gauge-invariance, \cite {SW}. Three additional
gauge-invariant variables, which play a crucial role in our analysis, are the
divergence of the acceleration (i.e. $A\equiv\nabla_ia^i$), its spatial
gradient (i.e. $A_i\equiv\mbox{}^{(3)}\nabla_iA$), and the spatial gradient of
the curvature scalar $K$ associated with $\Sigma_{\perp}$ (i.e.
$K_i\equiv\mbox{}^{(3)}\nabla_iK$).

It is convenient to introduce the following local decomposition for the
spatial gradient of $D_i$, \cite{EBH},
\begin{equation}
\Delta_{ij}\equiv S\mbox{}^{(3)}\nabla_jD_i=W_{ij}+\Sigma_{ij}+ \frac{1}{3}%
\Delta h_{ij} ,  \label{Dgr}
\end{equation}
where $W_{ij}\equiv\Delta_{[ij]}$ contains information about the rotational
behaviour of $D_i$,
$\Sigma_{ij}\equiv\Delta_{(ij)}-\Delta_i^{\hspace{1mm}i}h_{ij}/3$ describes
the formation of anisotropies (e.g. pancakes or cigar-like structures), and
$\Delta\equiv\Delta_i^{\hspace{1mm}i}$ is related to the spherically symmetric
gravitational clumping of matter.  Although in a general perturbation pattern
we expect turbulence (i.e. $W_{ij}\neq0$) and anisotropic structures (i.e.
$\Sigma_{ij}\neq0$ as well as material aggregation (i.e. $\Delta>0$), it is
the latter scalar which is crucial for the structure formation purposes.

\section{The Linear Regime}
%%%%%%%%%%%%%%%%%%%%%%%%%%%
In reality $\mu$, $p$, $\Theta$ and $H_i$ must have a spatial dependence as
well as a temporal one. Moreover, $K$, $a_i$, $\sigma_{ij}$, $\omega_{ij}$,
$\Pi_{ij}$, $E_{ij}$ and $H_{ij}$ will generally take non-zero
values. Assuming that the observed universe is close to a FFRW spacetime, we
can linearize the evolution equations by treating all the gauge-invariant
quantities, along with their derivatives, as first-order variables. The exact,
fully non-linear formulae have already been derived in \cite{TB}.  Here we
give their linearized versions only.

\subsection{Evolution Equations}
%%%%%%%%%%%%%%%%%%%%%%%%%%%%%%%%
The linear regime is monitored through the following combination of propagation
formulae and constraint equations:

{\bf (i)} The conservation laws of the energy and the momentum densities of the
fluid, respectively expressed by
\begin{equation}
\frac{\dot{\mu}}{\mu}+(1+w)\Theta=0 ,  \label{ledc}
\end{equation}
and
\begin{equation}
\kappa\mu(1+w)a_i+Y_i-\frac{2}{S}{\cal M}_{[ij]}H^j=0 ,  \label{lmdc}
\end{equation}
where the ratio $w\equiv p/\mu c^2$ evolves according to
\begin{equation}
\dot{w}=-(1+w)\left(\frac{c_s^2}{c^2}-w\right)\Theta ,  \label{wprop}
\end{equation}
with $c_s^2\equiv\dot{p}/\dot{\mu}$ representing the adiabatic
sound speed.  Although generally $w$ is allowed to vary, when it remains
constant along the fluid-flow lines (i.e. when $\dot{w}=0$) equation
(\ref{wprop}) suggests that $w=c_s^2/c^2=constant$, provided of course that
$\Theta\neq0$.

{\bf (ii)} The propagation equations that determine the kinematics of the
universe.  These are Raychaudhuri's formula,
\begin{equation}
\dot{\Theta}+\frac{\Theta^2}{3}+\frac{\kappa\mu c^4}{2}\left(1+3w\right)-
A-\Lambda c^2=0 ,  \label{lRay}
\end{equation}
where $\Lambda$ is the cosmological constant, and the propagation formulae of
the vorticity\footnote{Equation (\ref{lvprop}) monitors the model's rotational
behaviour through the vorticity tensor, as opposed to the vorticity vector used
in \cite{TB} (see eqn (89) there).  Recalling that
$\omega_i=\eta_{ijkq}u^j\omega^{kq}/2c$ the equivalence of the two formulae
becomes evident.} and the shear tensors, respectively given by
\begin{equation}
\dot{\omega}_{ij}+\frac{2\Theta}{3}\omega_{ij}=
\mbox{}^{(3)}\nabla_{[j}a_{i]} ,  \label{lvprop}
\end{equation}
and
\begin{equation}
\dot{\sigma}_{ij}+\frac{2\Theta}{3}\sigma_{ij}=
\mbox{}^{(3)}\nabla_{(j}a_{i)}- \frac{A}{3}h_{ij}+
\frac{\kappa c^2}{2}\Pi_{ij}-c^2E_{ij} ,  \label{lshprop}
\end{equation}
where $E_{ij}\equiv C_{ikjq}u^ku^q/c^2$ is the ``electric'' part of the Weyl
tensor.  To first order, the scalar $A$ is given by the 3-divergence of the
fluid acceleration (i.e $A=\mbox{}^{(3)}\nabla^ia_i$).

{\bf (iii)} The linear propagation equations of $D_i$, ${\cal Z}_i$ and
${\cal M}_{ij}$, respectively governing the growth of spatial inhomogeneities
in the energy density of the fluid,
\begin{equation}
\dot{D}_i=w\Theta D_i-\left(1+w\right){\cal Z}_i-
\frac{2\Theta}{\kappa\mu c^2}{\cal M}_{[ij]}H^j+
\frac{2S\Theta H^2}{3\mu c^4}a_i ,  \label{lDprop}
\end{equation}
in the expansion scalar,\footnote{In \cite{TB}, based on the weakness of the
magnetic field, we ignored the linear effects on the evolution of the expansion
and the 3-curvature gradients resulting from the field's contribution to the
active gravitational mass of the universe. Here we fully incorporate these
effects via the second last terms in the right hand side of (\ref{lcZprop})
and (\ref{lCi}) (compare them to eqns (91) and (99) in \cite{TB}). Notice that
these quantities provide all the coupling between the magnetic and the matter
inhomogeneities that is left, once the infinite conductivity approximation is
abandoned in favour of a pure source-free magnetic field (see appendix B in
\cite{TB}). Although they make no qualitative difference and introduce
negligible quantitative changes, both terms are included here for completeness.}
\begin{equation}
\dot{{\cal Z}}_i=-\frac{2\Theta}{3}{\cal Z}_i-
\frac{\kappa\mu c^4}{2}D_i-3c^2{\cal M}_{[ij]}H^j-
c^2{\cal M}_{ji}H^j+SA_i ,  \label{lcZprop}
\end{equation}
and in the magnetic field vector,
\begin{eqnarray}
\dot{{\cal M}}_{ij}&=&-\frac{2\Theta}{3}{\cal M}_{ij}-
\frac{2\kappa}{3}H_i{\cal Z}_j+
\kappa SH^k\mbox{}^{(3)}\nabla_j\left(\sigma_{ik}+\omega_{ik}\right)-
\frac{\kappa\Theta S}{3c^2}\left(2H_ia_j+a_iH_j\right)+\nonumber\\
&\mbox{}&\frac{\kappa\Theta S}{3c^2}a_kH^kh_{ij}+
\kappa h_i^{\hspace{1mm}k}R_{kqjs}H^qu^s ,  \label{lcMprop}
\end{eqnarray}
recalling that $A_i=\mbox{}^{(3)}\nabla_iA$ by definition.

{\bf (iv)} The evolution of the magnetic field is governed by the four
decomposed Maxwell's equations (see eqns (\ref{M1})-(\ref{M4})), of which only
\begin{equation}
\nabla_iH^i=\frac{1}{c^2}a_iH^i ,  \label{lM3}
\end{equation}
and
\begin{equation}
h_i^{\hspace{1mm}j}\dot{H}_j=
\left(\sigma^i_{\hspace{1mm}j}+\omega^i_{\hspace{1mm}j}
-\frac{2\Theta}{3}h^i_{\hspace{1mm}j}\right)H^j ,  \label{lM4}
\end{equation}
are crucial for our analysis.  The former verifies that the magnetic field is
a ``solenoidal'' (i.e. $\mbox{}^{(3)}\nabla_iH^i=0$), and the latter, when
contracted with the field vector, provides a radiation-like linear
evolution law for the magnetic energy density
\begin{equation}
H^2=\frac{\Bbb{H}}{S^4} ,  \label{lH2ev}
\end{equation}
where $\dot{\Bbb{H}}=0$.

{\bf (v)} We close this section with a brief discussion on the geometry of
$\Sigma_{\perp}$, the observer's instantaneous rest space.  Its curvature is
characterized by the scalar
\begin{equation}
K=2\left(\kappa\mu c^2-\frac{\Theta^2}{3c^2}+\Lambda\right) ,
\label{lK}
\end{equation}
so that $\Theta^2/3=\kappa\mu c^4+\Lambda c^2$ to zero-order.  When there is
no vorticity and only then, $K$ coincides with the 3-Ricci scalar of
the spacelike hypersurfaces that define the instantaneous rest space of all
the fundamental observers.  Its propagation formula,
\begin{equation}
\dot{K}=-\frac{2\Theta}{3}\left(K+\frac{2}{c^2}A\right) ,
\label{lKprop}
\end{equation}
suggests that in the linear regime the fluid acceleration acts as the sole
source of spatial curvature through its 3-divergence.  Following \cite{EBH},
we describe the spatial variations of the 3-curvature by the gauge-invariant
vector $C_i\equiv S^3K_i$ and provide a supplementary relation between
$D_i$, $Z_i$ and ${\cal M}_{ij}$,
\begin{equation}
C_i=2\kappa\mu c^2S^2D_i+2S^2{\cal M}_{ji}H^j-
\frac{4\Theta S^2}{3c^2}{\cal Z}_i ,  \label{lCi}
\end{equation}
which, by means of (\ref{lDprop})-(\ref{lcMprop}), leads to
\begin{equation}
\dot{C}_i=-\frac{4\Theta S^3}{3c^2}A_i .  \label{lCprop}
\end{equation}
The above propagation formula is consistent with equation (\ref{lKprop}) and,
together with (\ref{lbAi}), confirms the interdependence between the spatial
curvature and the acceleration of the fluid flow.  The advantages of choosing
$C_i$, instead of $K_i$, to describe the spatial variations in the 3-curvature,
will become clear later.

\section{The Case of a Barotropic Perfect Fluid}
%%%%%%%%%%%%%%%%%%%%%%%%%%%%%%%%%%%%%%%%%%%%%%%%
Among the propagation formulae given above, which refer to a general perfect
fluid with pressure, there is no equation for the evolution of 3-gradients in
the pressure. The reason is that the propagation of $Y_i$ will be determined
directly from (\ref{lDprop}), once the material content of the universe has
been specified.

\subsection{Equation of State}
%%%%%%%%%%%%%%%%%%%%%%%%%%%%%%
Here, we extend the analysis presented in~\cite{TB} by considering a universe
filled with a single barotropic perfect fluid. Its equation of state
is\footnote{From now on, all our results will refer to a barotropic fluid
unless otherwise stated. We will also ignore the entropy contribution to the
fluid-pressure.}
\begin{equation}
p=p(\mu) ,  \label{eos}
\end{equation}
suggesting that $\nabla_{[i}p\nabla_{j]}\mu=0$. Consequently, the relation
between pressure and energy density gradients becomes
\begin{equation}
SY_i=\kappa\mu c_s^2D_i ,  \label{YiDi}
\end{equation}
since $c_s^2=dp/d\mu$ relative to the observer's rest frame.

\subsection{Kinematic Evolution}
%%%%%%%%%%%%%%%%%%%%%%%%%%%%%%%%
\subsubsection{The Acceleration}
%%%%%%%%%%%%%%%%%%%%%%%%%%%%%%%%
The energy density conservation law of the barotropic fluid is still
expressed by (\ref{ledc}). However, the momentum density conservation law,
(\ref{lmdc}), together with (\ref{YiDi}), gives
\begin{equation}
a_i=\frac{1}{(1+w)S}\left(\frac{2}{\kappa\mu}{\cal M}_{[ij]}H^j-
c_s^2D_i\right) ,  \label{ai}
\end{equation}
for the acceleration of a fundamental observer. It depends both on gradients
in the energy density of the fluid and on gradients in the magnetic
field. Thus, the geodesic flow can still be preserved provided that the field
gradients counterbalance those of the material component. The necessary and
sufficient condition for this to occur is\footnote{The geodesic flow condition
can simplify the evolutionary relations of section 3.2 considerably. However,
it does not appear to be consistent and we will not pursue the matter any
further here.} $D_i=2{\cal M}_{[ij]}H^j/\kappa\mu c_s^2$.

Unlike the pressure-free case (see \cite{TB}), the acceleration of the
barotropic fluid is not always normal to the magnetic-field
vector.  Alternatively, one might say that, when $p\neq0$, the time derivative
$\dot{H}_i$ no longer lies on $\Sigma_{\perp}$.  We verify these statements by
simply contracting (\ref{ai}) with $H_i$.  We find that
\begin{equation}
a_iH^i=u_i\dot{H}^i=-\frac{c_s^2}{(1+w)S}H^iD_i ,  \label{aiHi}
\end{equation}
where generally $H_iD^i\neq0$. Obviously, $a_i$ and $H_i$ remain orthogonal if
$H_iD^i=0$. We can modify this condition by setting $E\equiv H^2/\mu c^2$,
taking its 3-gradient and then contracting with the magnetic field vector. The
result,
\begin{equation}
H^iD_i=-\frac{S}{E}H^i\mbox{}^{(3)}\nabla_iE ,  \label{ocaiHi}
\end{equation}
suggests that the acceleration of the fluid flow remains normal to the magnetic
field if and only if the directional derivative $H^i\mbox{}^{(3)}\nabla_iE$
vanishes (i.e. when the energy density ratio $E$ does not change along the
magnetic field lines). In this case, the time derivative of the magnetic field
lies on the observer's instantaneous rest space.  This fact can simplify, among
other, calculations involving commutations between the spatial gradients of
$\dot{H}_i$.

Spatial gradients in the fluid acceleration affect the expansion dynamics
directly (see eqns (\ref{lRay}), (\ref{lvprop}) and (\ref{lshprop})), as well
as the spatial geometry (see eqn (\ref{lKprop})).  Consequently the following
new decomposition of the acceleration's 3-gradient is of major importance. It
is obtained directly from equation (\ref{ai}) via the commutation laws for the
3-gradients of scalars and spacelike vectors (see eqns (\ref{scmtr}) and
(\ref{vcmtr}) in Appendix B), relations (\ref{Dgr}), (\ref{lM4}) and the
relativistic expression $\Pi_{ij}=H^2h_{ij}/3-H_iH_j$ for the magnetic
anisotropic stresses (see \S 7.4.1 in \cite{T} for more details).
\begin{eqnarray}
\mbox{}^{(3)}\nabla_ja_i&=&
-\frac{c_s^2}{(1+w)S^2}\left(\Sigma_{ij}+W_{ij}+\frac{\Delta}{3}h_{ij}\right)-
\frac{4\Theta H^2}{9\mu c^2(1+w)}\omega_{ij}+
\frac{H^2}{3\mu(1+w)}\mbox{}^{(3)}R_{ij}-\nonumber\\&\mbox{}&
\frac{1}{2\mu(1+w)}\mbox{}^{(3)}\nabla_j\mbox{}^{(3)}\nabla_iH^2+ 
\frac{1}{\kappa\mu(1+w)S}H^k\mbox{}^{(3)}\nabla_k{\cal M}_{ij} ,
\label{lbai3gr}
\end{eqnarray}
where $\mbox{}^{(3)}R_{ij}$ is the 3-Ricci tensor of the spacelike regions
(given by eqn (83) in \cite{TB}). In what follows, the trace, the skew part
and the symmetric part of the above will be employed to analyze the
magnetohydrodynamical effects upon the kinematics and the spatial geometry
of our cosmological model.

\subsubsection{The Deceleration Parameter}
%%%%%%%%%%%%%%%%%%%%%%%%%%%%%%%%%%%%%%%%%%
To begin with, the trace of (\ref{lbai3gr}),
\begin{equation}
A=\mbox{}^{(3)}\nabla^ia_i=-\frac{c_s^2}{(1+w)S^2}\Delta+
\frac{H^2}{3\mu(1+w)}K-\frac{H^2}{2\mu(1+w)S^2}{\cal B},  \label{lbA}
\end{equation}
where ${\cal B}\equiv S^2\mbox{}^{(3)}\nabla^2H^2/H^2$ , is substituted into
(\ref{lRay}) to produce Raychaudhuri's equation for a magnetized universe
filled with a single barotropic perfect fluid of infinite conductivity. This
formula is recast into the following alternative expression
\begin{equation}
\frac{\Theta^2}{3c^2}q=\frac{\kappa\mu c^2}{2}(1+3w)-
\frac{H^2}{3\mu c^2(1+w)}K+
\frac{1}{(1+w)S^2}\left(\frac{c_s^2}{c^2}\Delta
+\frac{H^2}{2\mu c^2}{\cal B}\right)-
\Lambda ,  \label{lbdec}
\end{equation}
where $q\equiv-\ddot{S}S/\dot{S}^2$ is the ``deceleration parameter'' (note
that $\Delta=S^2\mbox{}^{(3)}\nabla^2\mu/\mu$ to first order). Clearly,
the sign of the quantity on the right hand side of (\ref{lbdec}) determines
whether the expansion slows down or continues unimpeded. Not surprisingly, a
spherically symmetric increase in the energy density of the field (i.e.
${\cal B}>0$), together with any material aggregation (i.e. $\Delta>0$), slows
the expansion down.  Their combined effect is of the first order and, as the
Laplacians verify, it is confined to regions well within the horizon.  However,
while the energy density of ordinary matter (i.e. $w>-1/3$) always adds a
positive value to the deceleration parameter, the contribution of the magnetic
energy density depends on the geometry of $\Sigma_{\perp}$.  According to
(\ref{lbdec}), the coupling between the field and the 3-curvature slows down
the expansion of spatially open almost-FRW universe (i.e. when $K<0$) but
accelerates perturbed Friedmannian cosmologies with positive spatial curvature
(i.e. $K>0$).  This rather unconventional magnetic effect is global, though
still first order in magnitude since $K=0$ in the background.  It vanishes
when the perturbed universe retains its spatial flatness.

\subsubsection{The Vorticity Tensor}
%%%%%%%%%%%%%%%%%%%%%%%%%%%%%%%%%%%%
According to (\ref{lvprop}), only the antisymmetric part of (\ref{lbai3gr})
affects the vorticity propagation.  Since
$W_{ij}=-(1+w)\Theta S^2\omega_{ij}/c^2$ to the first order (see \cite{EBH})
and $\mbox{}^{(3)}\nabla_{[i}\mbox{}^{(3)}\nabla_{j]}H^2=
4\Theta H^2\omega_{ij}/3c^2$, as the commutator of the 3-gradients of scalars
(see eqn (\ref{scmtr}) in Appendix B) and the last of Maxwell's equations (see
eqn (\ref{lM4})) imply, we obtain
\begin{equation}
\dot{\omega}_{ij}+
\frac{2\Theta}{3}\left(1-\frac{3c_s^2}{2c^2}\right)\omega_{ij}=
\frac{1}{\kappa\mu(1+w)S}H^k\mbox{}^{(3)}\nabla_k{\cal M}_{[ij]} .
\label{lbvprop}
\end{equation}
Notice that a cosmic magnetic field influences the vorticity of the universe
solely through the antisymmetric part of the gradient field ${\cal M}_{ij}$,
which itself describes the rotational behaviour of the magnetic field
vector (see Appendix C.1 in \cite{TB}).  Also, according to (\ref{lbvprop}),
the field has no effect at all when the directional derivative
$H^k\mbox{}^{(3)}\nabla_k{\cal M}_{[ij]}$ vanishes, that is when
$\mbox{\rm curl}H_i$ does not change along the magnetic field lines.

\subsubsection{The Shear Tensor}
%%%%%%%%%%%%%%%%%%%%%%%%%%%%%%%%
The symmetric part of (\ref{lbai3gr}) together with its trace allows us to
recast equation (\ref{lshprop}), for the linear evolution of the shear tensor,
into the following form
\begin{eqnarray}
\dot{\sigma}_{ij}+\frac{2\Theta}{3}\sigma_{ij}&=&
-\frac{c_s^2}{(1+w)S^2}\Sigma_{ij}-
\frac{1}{2\mu(1+w)}\left(\mbox{}^{(3)}\nabla_{(i}\mbox{}^{(3)}\nabla_{j)}
-\frac{1}{3}h_{ij}\mbox{}^{(3)}\nabla^2\right)H^2+\nonumber\\&\mbox{}&
\frac{H^2}{3\mu(1+w)}\left(\mbox{}^{(3)}R_{(ij)}-\frac{{\cal K}}{3}h_{ij}\right)+
\frac{\kappa c^2}{2}\Pi_{ij}+\nonumber\\&\mbox{}&
\frac{1}{\kappa\mu(1+w)S}H^k\mbox{}^{(3)}\nabla_k{\cal M}_{(ij)}-
c^2E_{ij} .
\label{lbshprop}
\end{eqnarray}
Clearly, the shear anisotropies evolve in a rather complicated way under the
simultaneous influence of a number of sources. According to (\ref{lbshprop})
such sources are: the fluid; the magnetic field; the geometry of the observer's
3-D rest space; and the long range source-free gravitational field.  The
magnetic influence is multi faceted.  In particular, anisotropic spatial
variations in the energy density of the field have a similar effect to those
in the energy density of the fluid (the latter represented by
$\Sigma_{ij}$).  Also, the magnetic energy density couples with anisotropies
in the spatial curvature to create an additional effect.  Notice that any
anisotropic patterns in the distribution of the magnetic field vector
(described by ${\cal M}_{(ij)}$ since ${\cal M}_i^{\hspace{1mm}i}=0$) exert
no influence at all if the directional derivative
$H^k\mbox{}^{(3)}\nabla_k{\cal M}_{(ij)}$ vanishes.

\subsubsection{The 3-curvature Scalar}
%%%%%%%%%%%%%%%%%%%%%%%%%%%%%%%%%%%%%%
In the linear regime, the curvature scalar of the observer's instantaneous
3-D rest space evolves according to equation (\ref{lKprop}).  Substituting
the trace of (\ref{lbai3gr}) into the latter we obtain
\begin{equation}
\dot{K}+\frac{2\Theta}{3}\left(1+\frac{2H^2}{3\mu c^2(1+w)}\right)K=
\frac{4\Theta}{3(1+w)S^2}
\left(\frac{c_s^2}{c^2}\Delta+\frac{H^2}{2\mu c^2}{\cal B}\right) ,
\label{lbKprop}
\end{equation}
or, since $H^2/\mu c^2(1+w)\ll1$,
\begin{equation}
\dot{K}+\frac{2\Theta}{3}K=\frac{4\Theta}{3(1+w)S^2}
\left(\frac{c_s^2}{c^2}\Delta+\frac{H^2}{2\mu c^2}{\cal B}\right) .
\label{lbKprop1}
\end{equation}
Therefore, on regions of subhorizon size, any spherically symmetric spatial
increase in the energy density of the fluid (i.e $\Delta>0$), or of the
magnetic field (i.e. ${\cal B}>0$), acts as a source of positive
curvature. This is a first order effect similar to the small-scale
magnetohydrodynamical impact on the expansion of the universe (see the last
term in the right hand side of eqn (\ref{lbdec})). The magnetic influence also
results in a global effect of the opposite type. As (\ref{lbKprop}) reveals,
the field tends to smooth out the curvature of the spacelike regions through
its coupling with the background expansion. This ``magnetic smoothing'', which
here is of negligible magnitude, is analogous to the field's global
magneto-geometrical impact upon the universal expansion (compare to the second
term in the right hand side of (\ref{lbdec})) both qualitatively and
quantitatively.

We close this section with a sort comment on the non-local magnetic effects
illustrated in equations (\ref{lbdec}) and (\ref{lbKprop}). We attribute such
behaviour to the vectorial nature of the field, as opposed to the scalar nature
of quantities such as the energy density of the fluid or its isotropic
pressure. Being a vector, the field interacts with the curvature of the
spacelike regions (e.g. through the 3-Ricci identity) and this dependence
creates the aforementioned effects. However, we do not suggest that any
perturbed spacelike vector would have a similar impact. The magnetic influence
outlined above depends crucially upon the specific properties of the field, as
these are reflected in Maxwell's equations, and also in the unique way general
relativity describes the magnetic anisotropic stresses (see comments on the
derivation of eqn (\ref{lbai3gr})).

\subsection{Dynamic Evolution}
%%%%%%%%%%%%%%%%%%%%%%%%%%%%%%
\subsubsection{The Growth of the Inhomogeneities}
%%%%%%%%%%%%%%%%%%%%%%%%%%%%%%%%%%%%%%%%%%%%%%%%%
The introduction of a barotropic fluid modifies the key linear propagation
equations (\ref{lDprop}), (\ref{lcZprop}), (\ref{lcMprop}) and
(\ref{lCprop}). More specifically, by means of (\ref{ai}), the former becomes
\begin{equation}
\dot{D}_i=w\Theta D_i-(1+w){\cal Z}_i-
\frac{2\Theta}{\kappa\mu c^2}{\cal M}_{[ij]}H^j .  \label{blDprop}
\end{equation}
The direct barotropic influence on the evolution of the expansion gradients
comes through the spatial gradient of $A$,\footnote{In deriving (\ref{lbAi})
we have treated $\mbox{}^{(3)}\nabla_iw$ and $\mbox{}^{(3)}\nabla_ic_s^2$ as
first order gauge-invariant quantities. Though the former is straightforward to
prove, the latter requires the gauge-independence of
$\mbox{}^{(3)}\nabla_i\dot{p}$ and $\mbox{}^{(3)}\nabla_i\dot{\mu}$ to be shown
first.}
\begin{equation}
A_i=-\frac{c_s^2}{(1+w)S}\mbox{}^{(3)}\nabla^2D_i+\frac{H^2}{3\mu(1+w)S^3}C_i-
\frac{1}{2\mu(1+w)}\mbox{}^{(3)}\nabla^2{\cal H}_i-
\frac{2c_s^2\Theta}{c^2}\mbox{}^{(3)}\nabla_j\omega_j^{\hspace{1mm}j},
\label{lbAi}
\end{equation}
where ${\cal H}_i\equiv\mbox{}^{(3)}\nabla_iH^2$.  Substituting the above into
(\ref{lcZprop}) and using (\ref{lCi}) we obtain
\begin{eqnarray}
\dot{{\cal Z}}_i&=&-\frac{2\Theta}{3}{\cal Z}_i-\frac{\kappa\mu c^4}{2}D_i-
3c^2{\cal M}_{[ij]}H^j-c^2{\cal M}_{ji}H^j-
\frac{c_s^2}{1+w}\mbox{}^{(3)}\nabla^2D_i-\nonumber\\
&\mbox{}&\frac{S}{2\mu(1+w)}\mbox{}^{(3)}\nabla^2{\cal H}_i-
\frac{2c_s^2\Theta S}{c^2}\mbox{}^{(3)}\nabla_j\omega_i^{\hspace{1mm}j} ,
\label{blcZprop}
\end{eqnarray}
By means of (\ref{Wt}) and the fact that $h_i^{\hspace{1mm}j}R_{jk}u^k=0$ (see
eqn (63) in \cite{TB}), the last term in (\ref{lcMprop}), which describes the
effects of spacetime curvature upon the evolution of magnetic inhomogeneities,
becomes
\begin{equation}
\kappa h_i^{\hspace{1mm}k}R_{kqjs}H^qu^s=
\kappa h_i^{\hspace{1mm}k}C_{kqjs}H^qu^s .  \label{ct}
\end{equation}
Moreover, using decomposition (\ref{dWt}) of the Weyl tensor we may recast the
above into
\begin{equation}
\kappa h_i^{\hspace{1mm}k}C_{kqjs}H^qu^s=
-\kappa\eta_{ik}^{\hspace{2mm}qs}H^ku_qH_{sj} ,  \label{ct1}
\end{equation}
showing that only spacetime ripples caused by the long-range gravitational
forces, here represented by the magnetic part
$H_{ij}\equiv\eta_{ip}^{\hspace{2mm}kq}C_{kqjs}u^pu^s/2c^2$ of the Weyl tensor,
affect the propagation of spatial inhomogeneities in the cosmic magnetic
field. The above result together with equations (\ref{ai}) and (\ref{aiHi})
allows us to transform (\ref{lcMprop}) into
\begin{eqnarray}
\dot{{\cal M}}_{ij}&=&-\frac{2\Theta}{3}{\cal M}_{ij}-
\frac{2\kappa}{3}H_i{\cal Z}_j+
\kappa SH^k\mbox{}^{(3)}\nabla_j\left(\sigma_{ik}+\omega_{ik}\right)+
\frac{2\Theta H^2}{9\mu c^2(1+w)}{\cal M}_{[ij]}+\nonumber\\&\mbox{}&
\frac{\kappa c_s^2\Theta}{3c^2(1+w)}\left(2H_iD_j+H_jD_i-H^kD_kh_{ij}\right)+
\kappa S\eta_{ik}^{\hspace{2mm}qs}H^ku_qH_{sj} .  \label{blcMprop}
\end{eqnarray}
According to (\ref{blDprop}) it is the only the contraction ${\cal M}_{[ij]}H^j$
that contributes to the linear growth of spatial inhomogeneities in the energy
density of the medium. Therefore, taking the skew part of (\ref{blcMprop}) and
then contracting with the magnetic field vector, we obtain\footnote{See \S 5.4
and \S 7.5 in \cite{T} for a detailed derivation of the complete set of the
exact and linear propagation equations.}
\begin{eqnarray}
\dot{{\cal M}}_{[ij]}H^j&=&-\frac{2\Theta}{3}{\cal M}_{[ij]}H^j-
\frac{2\kappa}{3}H_{[i}{\cal Z}_{j]}H^j+
\kappa S h_{[i}^{\hspace{2mm}k}h_{j]}^{\hspace{1mm}q}
\left(\sigma_{ks}+\omega_{ks}\right)_{;q}H^sH^j+\nonumber\\&\mbox{}&
\frac{\kappa c_s^2\Theta}{3c^2(1+w)}H_{[i}D_{j]}H^j ,  \label{skewM}
\end{eqnarray}
which will prove useful later. Notice the absence of the spacetime curvature
term from the right hand side of equation (\ref{skewM}). As a result, long
range gravitational forces have no linear effect on the evolution of magnetized
density perturbations.

As far as spatial inhomogeneities in the curvature scalar are concerned,
equations (\ref{YiDi}), (\ref{ai}) and (\ref{lbAi}) reshape their propagation
formula, (\ref{lCprop}), into the following
\begin{equation}
\dot{C}_i=\frac{4c_s^2\Theta S^2}{3c^2(1+w)}\mbox{}^{(3)}\nabla^2D_i+
\frac{2\Theta S^3}{3\mu c^2(1+w)}\mbox{}^{(3)}\nabla^2{\cal H}_i+
\frac{8c_s^2\Theta^2S^3}{3c^4}\mbox{}^{(3)}\nabla_j\omega_i^{\hspace{1mm}j} ,
\label{blCprop}
\end{equation}
implying that the gradient field $C_i$ is invariant on large scales (i.e. when
the Laplacian terms are negligible) if
$\mbox{}^{(3)}\nabla_j\omega_i^{\hspace{1mm}j}=0$. Finally, under the
barotropic fluid assumption, reflected in (\ref{aiHi}), Maxwell's equations
become,
\begin{equation}
\nabla_iH^i= -\frac{c_s^2}{c^2(1+w)S}H^iD_i  \label{blM3}
\end{equation}
and
\begin{equation}
\dot{H}_i= \left(\sigma_{ij}+ \omega_{ij}- \frac{2\Theta}{3}%
h_{ij}\right)H^j- \frac{c_s^2}{c^2(1+w)S}H^jD_ju_i ,  \label{blM4}
\end{equation}
with the non-zero right hand side of (\ref{blM3}) and the last term of
(\ref{blM4}) being direct results of the changes in the fluid motion relative
to the pressureless case.

\subsubsection{The Growth of the Density Gradient}
%%%%%%%%%%%%%%%%%%%%%%%%%%%%%%%%%%%%%%%%%%%%%%%%%%
The dynamics of the inhomogeneity variable $D_i$ is governed by (\ref{blDprop}),
together with equations (\ref{blcZprop})\footnote{Alternatively, one can use
(\ref{blCprop}) instead of (\ref{blcZprop}), on substituting ${\cal Z}_i$ by
$C_i$ in (\ref{blDprop}) from (\ref{lCi}).} and (\ref{skewM}), or by the linear
second order differential equation, which follows from (\ref{blDprop}), by
means of (\ref{ledc}), (\ref{wprop}), (\ref{lRay}), (\ref{blcZprop}),
(\ref{skewM}) and (\ref{blM4}). This equation is
\begin{eqnarray}
\ddot{D}_i&=&-\left(\frac{2}{3}+\frac{c_s^2}{c^2}-2w\right)\Theta\dot{D}_i+ 
\nonumber\\&\mbox{}&
\left(\left(\frac{1}{2}-\frac{3c_s^2}{c^2}+4w-\frac{3w^2}{2}\right)\kappa\mu c^4
-\left(\frac{3c_s^2}{c^2}-5w\right)\Lambda c^2\right)D_i+
\nonumber\\&\mbox{}&
c_s^2\mbox{}^{(3)}\nabla^2D_i+
\nonumber\\&\mbox{}&
\frac{2c_s^2\Theta S(1+w)}{c^2}\mbox{}^{(3)}\nabla_j\omega_i^{\hspace{1mm}j}-
\nonumber\\&\mbox{}&
\left(\left(\frac{c_s^2}{c^2}-w\right)6c^2
+\left(1+\frac{c_s^2}{c^2}\right)\frac{6\Lambda}{\kappa\mu}\right){\cal M}_{[ij]}H^j+
\nonumber\\&\mbox{}&
\frac{S}{2\mu}\mbox{}^{(3)}\nabla^2{\cal H}_i-
\nonumber\\&\mbox{}&
\frac{2\Theta S}{\mu c^2}\mbox{}^{(3)}\nabla_{[j}\dot{H}_{i]}H^j.
\label{lddD}
\end{eqnarray}
This is the generalization of formula (26) in~\cite{EBH} for a magnetized
almost-FFRW universe. It has the form of a wave equation with extra terms due
to the universal expansion, gravity, the cosmological constant, the magnetic
field and the vorticity. The difference in the vorticity terms between
equation (\ref{lddD}) above, and its corresponding formula (115) in \cite{TB},
is due to the residual coupling between the divergence of the vorticity tensor
and the energy density of the field that remains when $p=0$.

\section{The Scalar Variables}
%%%%%%%%%%%%%%%%%%%%%%%%%%%%%%
So far we have considered the evolution of gauge-invariant vector variables
and in particular the propagation of $D_i$, the gradient field that describes
orthogonal to the fluid flow variations of the energy density. However,
regarding the growth (or decay) of density inhomogeneities, the vector field
$D_i$ contains more information than actually required. We can extract the
information we need by adopting the local decomposition (\ref{Dgr}). Of the
three additional variables mentioned there, the scalar
$\Delta\equiv S\mbox{}^{(3)}\nabla^iD_i$ (alternatively
$\Delta=(S^2/\mu)\mbox{}^{(3)}\nabla^2\mu$ to first order) is the most
important one when addressing the problem of structure formation.

\subsection{Definitions}
%%%%%%%%%%%%%%%%%%%%%%%%
Focusing upon $\Delta$, which describes spherically symmetric spatial variations
in the energy density of the matter, we also consider the following
complementary scalar variables,
\begin{equation}
{\cal Z}\equiv S\mbox{}^{(3)}\nabla^i{\cal Z}_i ,\hspace{2cm}
{\cal B}\equiv\frac{S^2}{H^2}\mbox{}^{(3)}\nabla^2H^2 ,  \label{scZcB}
\end{equation}
respectively related to spatial gradients in the expansion and the energy
density of the magnetic field, and 
\begin{equation}
{\cal K}=S^2K ,  \label{Cal K}
\end{equation}
representing perturbations in the spatial curvature. Notice that all but
${\cal Z}$ are dimensionless variables. Also, ${\cal B}$ describes spherically
symmetric spatial variations in the energy density of the magnetic field and
it will be treated as the magnetic analogue of $\Delta$.

\subsection{Evolutionary Equations}
%%%%%%%%%%%%%%%%%%%%%%%%%%%%%%%%%%%
The propagation equations associated with the above defined scalars (see \S 7.6
in \cite{T} for details on their derivation) are
\begin{equation}
\dot{\Delta}=w\Theta\Delta-(1+w){\cal Z}-
\frac{\Theta H^2}{3\mu c^2}{\cal K}+\frac{\Theta H^2}{2\mu c^2}{\cal B} ,
\label{lbDelprop}
\end{equation}
\begin{eqnarray}
\dot{{\cal Z}}&=&-\frac{2\Theta}{3}{\cal Z}-\frac{\kappa\mu c^4}{2}\Delta-
\frac{c_s^2}{1+w}\mbox{}^{(3)}\nabla^2\Delta-
\frac{\kappa c^2H^2}{2}{\cal K}+\nonumber\\&\mbox{}&
\frac{\kappa c^2H^2}{4}{\cal B}-
\frac{H^2}{2\mu(1+w)}\mbox{}^{(3)}\nabla^2{\cal B} ,  \label{lbscZprop}
\end{eqnarray}
\begin{equation}
\dot{{\cal B}}=\frac{4c_s^2\Theta}{3c^2(1+w)}\Delta-\frac{4}{3}{\cal Z}-
\frac{4\Theta H^2}{9\mu c^2(1+w)}{\cal K} ,  \label{lbcBprop}
\end{equation}
and
\begin{equation}
\dot{{\cal K}}=\frac{4c_s^2\Theta}{3c^2(1+w)}\Delta+
\frac{2\Theta H^2}{3\mu c^2(1+w)}{\cal B} .  \label{lbcKprop}
\end{equation}
The first two are obtained by linearizing the 3-divergence of their
corresponding vector equations (\ref{blDprop}) and (\ref{blcZprop}). The third
results directly from definition (\ref{scZcB}b) via the laws governing
commutations between time derivatives and spatial gradients of scalars and
spacelike vectors (see eqns (\ref{scmtr1}) and (\ref{vcmtr1}) or eqn
(\ref{scmtr2}) in Appendix B). Finally, the last is a simple rearrangement of
(\ref{lbKprop1}).\footnote{The 3-divergence of (\ref{blCprop}) provides the
evolution formula of $C\equiv S\mbox{}^{(3)}\nabla^iC_i$, the scalar associated
with spatial inhomogeneities in the curvature of the spacelike regions. Its
form,
\begin{equation}
\dot{C}=\frac{4c_s^2\Theta S^2}{3c^2(1+w)}\mbox{}^{(3)}\nabla^2\Delta+
\frac{2\Theta S^2H^2}{3\mu c^2(1+w)}\mbox{}^{(3)}\nabla^2{\cal B} ,
\label{lbCprop}
\end{equation}
verifies that $C$ is time-invariant on large scales irrespective of the model's
rotational behaviour. Notice that one immediately recovers (\ref{lbcKprop}) from
(\ref{lbCprop}) on using definition (\ref{Cal K}) and the commutation law
(\ref{scmtr2}) in Appendix B.}

By combining equations (\ref{lbDelprop})-(\ref{lbcBprop}), or by linearizing
the 3-divergence of (\ref{lddD}), we obtain the following second order
differential equation for the evolution of the spatial matter aggregations
\begin{eqnarray}
\ddot{\Delta}&=&-\left(\frac{2}{3}+\frac{c_s^2}{c^2}-2w\right)
\Theta\dot{\Delta}+
\nonumber\\&\mbox{}&
\left(\left(\frac{1}{2}-\frac{3c_s^2}{c^2}+4w
-\frac{3w^2}{2}\right)\kappa\mu c^4-\left(\frac{3c_s^2}{c^2}
-5w\right)\Lambda c^2\right)\Delta+
\nonumber\\&\mbox{}&
c_s^2\mbox{}^{(3)}\nabla^2\Delta+
\nonumber\\&\mbox{}&
\left(\left(\frac{2}{3}-\frac{c_s^2}{c^2}+w\right)\kappa\mu c^2
-\left(\frac{1}{3}+\frac{c_s^2}{c^2}\right)\Lambda\right)
\frac{H^2}{\mu}{\cal K}-\nonumber\\&\mbox{}&
\left(\left(\frac{1}{2}-\frac{3c_s^2}{2c^2}+w\right)\kappa\mu c^2
-\left(\frac{1}{2}+\frac{3c_s^2}{2c^2}\right)\Lambda\right)
\frac{H^2}{\mu}{\cal B}+\nonumber\\&\mbox{}&
\frac{H^2}{2\mu}\mbox{}^{(3)}\nabla^2{\cal B} .  \label{lbddDel}
\end{eqnarray}
The rest of the variables evolve in accordance with the propagation formulae,
\begin{equation}
\dot{{\cal K}}=\frac{4c_s^2\Theta}{3c^2(1+w)}\Delta+
\frac{2\Theta H^2}{3\mu c^2(1+w)}{\cal B} ,  \label{lbdcK}
\end{equation}
and
\begin{equation}
\dot{{\cal B}}=\frac{4}{3(1+w)}\dot{\Delta}+
\frac{4\Theta}{3(1+w)}\left(\frac{c_s^2}{c^2}-w\right)\Delta ,
\label{lbdcB}
\end{equation}
where the latter is obtained by substituting ${\cal Z}$ in (\ref{lbcBprop})
from (\ref{lbDelprop}).\footnote{Equations (\ref{wprop}) and (\ref{lbdcB})
suggest that when $\dot{w}=0$, as it is the case in the dust era for example,
then $\dot{{\cal B}}=4\dot{\Delta}/3(1+w)$. So, during these periods spherically
symmetric spatial variations in the energy density of the magnetic field grow
(or decay) proportionally to those in the energy density of the matter.} This
is the system that governs the evolution of spatial matter aggregations in a
perturbed FFRW universe that contains a single barotropic perfect fluid of
infinite conductivity and is permeated by a weak cosmological magnetic field.

The equations obtained here are significantly simpler and more transparent than
their vector counterparts of section 4.3, especially as far as the role of the
magnetic field is concerned. The field no longer exerts its influence through
some complicated combinations of curl's and vector products, but simply via
the spatial gradients of its energy density. Moreover, these are exactly the
quantities that matter for structure formation purposes.

\section{Particular Solutions}
%%%%%%%%%%%%%%%%%%%%%%%%%%%%%%
In \cite{TB} we considered the evolution of density inhomogeneities during
the post-equilibrium era, when the universe is filled with a
non-relativistic perfect fluid (i.e. $p=0\Rightarrow w$, $c_s^2=0$). There,
based on the nature of the evolution equation for $\Delta$, we argued that the
magnetic effects on the growth of large-scale material aggregations are
relatively unimportant. Here, the existence of formulae (\ref{lbdcK}) and
(\ref{lbdcB}) will enable us to confirm, refine and extend these conclusions
as well as to study the behaviour of the density contrast in the radiation era.

\subsection{Harmonic Analysis}
%%%%%%%%%%%%%%%%%%%%%%%%%%%%%%
Following \cite{H}, \cite{EHB} and \cite{BDE}, we harmonically decompose the
inhomogeneity variable $\Delta$  by writing it in the form of the sum
\begin{equation}
\Delta=\sum_{n}\Delta^{(n)}Q^{(n)} ,  \label{nDel}
\end{equation}
with $\mbox{}^{(3)}\nabla_i\Delta^{(n)}=0$, $\dot{Q}^{(n)}=0$ and
$\mbox{}^{(3)}\nabla^2Q^{(n)}=-n^2Q^{(n)}/S^2$. Similarly, ${\cal K}$ and
${\cal B}$ may be written as
\begin{equation}
{\cal K}=\sum_{n}{\cal K}^{(n)}Q^{(n)} ,\hspace{1cm}\mbox{\rm and}\hspace{1cm}
{\cal B}=\sum_{n}{\cal B}^{(n)}Q^{(n)},  \label{ncKcB}
\end{equation}
where $\mbox{}^{(3)}\nabla_i{\cal K}=\mbox{}^{(3)}\nabla_i{\cal B}=0$. Notice
that the harmonic eigenvalue ($n$) coincides with the comoving wavenumber
($\nu$) because of the spatial flatness of the background universe.\footnote{If
the unperturbed universe has open spatial sections (i.e. $k=-1$) then
$n^2=\nu^2+1$, with $\nu^2\geq0$. Conversely, when the background model is
spatially closed (i.e. $k=+1$) the associated relation is $n^2=\nu(\nu+2)$,
where now $\nu=1,2,3,\ldots$, and the fundamental mode corresponds to $\nu=1$,
\cite{Ha}, \cite{DBE}.}

Substituting results (\ref{nDel}) and (\ref{ncKcB}) into equations
(\ref{lbddDel})-(\ref{lbdcB}) the harmonics decouple to provide the following
autonomous system
\begin{eqnarray}
\ddot{\Delta}^{(\nu)}&=&-\left(\frac{2}{3}+\frac{c_s^2}{c^2}-2w\right)
\Theta\dot{\Delta}^{(\nu)}+
\nonumber\\&\mbox{}&
\left(\left(\frac{1}{2}-\frac{3c_s^2}{c^2}+4w
-\frac{3w^2}{2}\right)\kappa\mu c^4-\frac{\nu^2c_s^2}{S^2}-
\left(\frac{3c_s^2}{c^2}-5w\right)\Lambda c^2\right)\Delta^{(\nu)}+
\nonumber\\&\mbox{}&
\left(\left(\frac{2}{3}-\frac{c_s^2}{c^2}+w\right)\kappa\mu c^2
-\left(\frac{1}{3}+\frac{c_s^2}{c^2}\right)\Lambda\right)c_A^2{\cal K}^{(\nu)}-
\nonumber\\&\mbox{}&
\left(\left(\frac{1}{2}-\frac{3c_s^2}{2c^2}+w\right)\kappa\mu c^2
+\frac{\nu^2}{2S^2}-\left(\frac{1}{2}+\frac{3c_s^2}{c^2}\right)\Lambda\right)
c_A^2{\cal B}^{(\nu)} ,  \label{nlbddDel}
\end{eqnarray}
\begin{equation}
\dot{{\cal K}}^{(\nu)}=\frac{4c_s^2\Theta}{3c^2(1+w)}\Delta^{(\nu)}+
\frac{2\Theta c_A^2}{3c^2(1+w)}{\cal B}^{(\nu)} ,  \label{nlbdcK}
\end{equation}
and
\begin{equation}
\dot{{\cal B}}^{(\nu)}=\frac{4}{3(1+w)}\dot{\Delta}^{(\nu)}+
\frac{4\Theta}{3(1+w)}\left(\frac{c_s^2}{c^2}-w\right)\Delta^{(\nu)} ,
\label{nlbdcB}
\end{equation}
where $c_A^2\equiv H^2/\mu$ is the Alfv\'{e}n speed characterizing the
propagation of hydromagnetic waves.

\subsection{The Radiation Era}
%%%%%%%%%%%%%%%%%%%%%%%%%%%%%%
When radiation dominates $w=c_s^2/c^2=1/3$ and the energy density of the
matter falls as $\mu=\Bbb{M}_R/S^4$ (see eqn (\ref{ledc})), suggesting,
together with equation (\ref{lH2ev}), that the the Alfv\'{e}n velocity remains
constant along the fluid-flow lines (i.e. $\dot{c}_A^2=0$). Ignoring the
cosmological constant (i.e. $\Lambda=0$), it is preferable to express
equations (\ref{nlbddDel})-(\ref{nlbdcB}) with respect to the scale factor,
$S(t)$,
\begin{equation}
S^2\frac{d^2\Delta^{(\nu)}}{dS^2}=
2\left(1-\frac{\nu^2S^2}{2\kappa\Bbb{M}_Rc^2}\right)\Delta^{(\nu)}+
\frac{2c_A^2}{c^2}{\cal K}^{(\nu)}-
\frac{c_A^2}{c^2}\left(1+\frac{3\nu^2S^2}{2\kappa\Bbb{M}_Rc^2}\right)
{\cal B}^{(\nu)} ,  \label{rnlbddDel}
\end{equation}
\begin{equation}
S\frac{d{\cal K}^{(\nu)}}{dS}=\Delta^{(\nu)}+
\frac{3c_A^2}{2c^2}{\cal B}^{(\nu)} ,  \label{rnlbdcK}
\end{equation}
\begin{equation}
\frac{d{\cal B}^{(\nu)}}{dS}=\frac{d\Delta^{(\nu)}}{dS} .  \label{rnlbdcB}
\end{equation}

In the long-wavelength limit (i.e. $\nu\rightarrow0$, or equivalently
$\nu^2S^2/\kappa\Bbb{M}_Rc^2\ll1$)\footnote{During the radiation epoch the
scale factor evolves as $S\equiv\beta t^{1/2}$, with
$\beta=(4\kappa\Bbb{M}_Rc^4/3)^{1/4}$. Considering a physical scale much
larger than the horizon (i.e. $\lambda_{phys}\gg d_H$), and taking into
account that $\lambda_{phys}\sim S\lambda_{com}$, $\lambda_{com}\sim1/\nu$ and
$d_H\sim ct$, we find that $\nu^2S^2/\kappa\Bbb{M}_Rc^2\ll1$ on large
scales. Clearly, subhorizon scales are characterized by the reverse inequality.}
the above system reduces to
\begin{equation}
S^2\frac{d^2\Delta^{(\nu)}}{dS^2}=2\Delta^{(\nu)}+
\frac{2c_A^2}{c^2}{\cal K}^{(\nu)}-
\frac{c_A^2}{c^2}{\cal B}^{(\nu)} ,  \label{LrnlbddDel}
\end{equation}
\begin{equation}
S\frac{d{\cal K}^{(\nu)}}{dS}=\Delta^{(\nu)}+
\frac{3c_A^2}{2c^2}{\cal B}^{(\nu)} ,  \label{LrnlbdcK}
\end{equation}
\begin{equation}
\frac{d{\cal B}^{(\nu)}}{dS}=\frac{d\Delta^{(\nu)}}{dS} .  \label{LrnlbdcB}
\end{equation}
and accepts a power-law solution of the form
\begin{equation}
\Delta^{(\nu)}(S)=\sum_{z}\Delta_z^{(\nu)}zS^z ,  \label{LrDel}
\end{equation}
where $\Delta_z^{(\nu)}$ are arbitrary positive constants. The parameter $z$
satisfies the cubic equation
\begin{equation}
z^3-z^2+\left(\frac{c_A^2}{c^2}-2\right)z-
\frac{2c_A^2}{c^2}\left(1+\frac{3c_A^2}{2c^2}\right)=0 ,  \label{ce2}
\end{equation}
which has three real roots provided that $c_A^2/c^2<3/11$, \cite{AS}, given in
trigonometric form by, \cite{Tu},
\begin{equation}
z\simeq\frac{1}{3}\left[1+2\sqrt{7}\left(1-\frac{3c_A^2}{14c^2}\right)
\cos\left(\frac{\theta+2k\pi}{3}\right)\right] ,  \label{rz}
\end{equation}
with $k=0$, $1$, $2$, and
\begin{equation}
\cos\theta\simeq\frac{10}{7\sqrt{7}}
\frac{1+\frac{9c_A^2}{4c^2}}{1-\frac{9c_A^2}{4c^2}} .  \label{rtheta}
\end{equation}
In the absence of a magnetic field (i.e. $c_A^2=0$), expressions (\ref{rz}),
(\ref{rtheta}) provide the standard solutions $z=0$,$2$, $-1$ associated with
a magnetic-free universe (see for example \cite{P} or \cite{EHB}).  However, the
coupling between the field and the 3-curvature obscures the overall magnetic
effect upon the growth of the density contrast.  As (\ref{rtheta}) reveals the
field increases the cosine term in (\ref{rz}) but at the same time decreases
this term's coefficient, with the net effect depending on the field's relative
strength.  To clarify the magnetic impact we consider the case of a spatially
flat (i.e. ${\cal K}^{(\nu)}=0$) perturbed universe. Then the system
(\ref{LrnlbddDel})-(\ref{LrnlbdcB}) reduces to
\begin{equation}
S^2\frac{d^2\Delta^{(\nu)}}{dS^2}=2\Delta^{(\nu)}-
\frac{c_A^2}{c^2}B^{(\nu)} ,  \label{LrnlbddDel1}
\end{equation}
\begin{equation}
\frac{dB^{(\nu)}}{dS}=\frac{d\Delta^{(\nu)}}{dS} ,  \label{LrnlbdcB1}
\end{equation}
while the density contrast evolves as
\begin{equation}
\Delta^{(\nu)}(S)=\Delta_1^{(\nu)}S^{z_1}+\Delta_2^{(\nu)}S^{z_2} ,
\label{LrDel1}
\end{equation}
where $\Delta_1^{(\nu)}$, $\Delta_2^{(\nu)}$ are constants and
\begin{equation}
z_{1,2}=\frac{1}{2}\left(1\pm3\sqrt{1-\frac{4c_A^2}{9c^2}}\right) ,
\label{Lz1}
\end{equation}
When the field is absent we recover the familiar evolution law (i.e.
$z_1=2$, $z_2=-1$) of a magnetic-free universe.  Generally however, the
large-scale magnetic effect is to reduce the growth rate of the density
contrast in proportion to its relative strength.

Conversely, the evolution of short-wavelength (i.e. $\nu\rightarrow\infty$, or
equivalently $\nu^2S^2/\kappa\Bbb{M}_Rc^2\gg1$) density aggregations is
governed by the following set of equations
\begin{equation}
S^2\frac{d^2\Delta^{(\nu)}}{dS^2}=
-\frac{\nu^2S^2}{\kappa\Bbb{M}_Rc^2}\Delta^{(\nu)}+
\frac{2c_A^2}{c^2}{\cal K}^{(\nu)}-
\frac{3\nu^2c_A^2S^2}{2\kappa\Bbb{M}_Rc^4}{\cal B}^{(\nu)} ,  \label{SrnlbddDel}
\end{equation}
\begin{equation}
S\frac{d{\cal K}^{(\nu)}}{dS}=\Delta^{(\nu)}+
\frac{3c_A^2}{2c^2}{\cal B}^{(\nu)} ,  \label{SrnlbdcK}
\end{equation}
\begin{equation}
\frac{d{\cal B}^{(\nu)}}{dS}=\frac{d\Delta^{(\nu)}}{dS} .  \label{SrnlbdcB}
\end{equation}
The lack of a general analytic solution forces us to ignore the effects of the
spatial curvature. In this case the remaining equations accept the solution
\begin{equation}
\Delta^{(\nu)}(S)=
\Delta_1^{(\nu)}\sin\left(\frac{\nu S}{c\sqrt{\kappa\Bbb{M}_R}}
\sqrt{1+\frac{3c_A^2}{2c^2}}\right)+
\Delta_2^{(\nu)}\cos\left(\frac{\nu S}{c\sqrt{\kappa\Bbb{M}_R}}
\sqrt{1+\frac{3c_A^2}{2c^2}}\right) .  \label{SrDel}
\end{equation}
So, small-scale matter aggregations oscillate with period
$2\pi c\sqrt{\kappa\Bbb{M}_R}/\nu\sqrt{1+3c_A^2/2c^2}$. Relative to the
non-magnetized case (see \cite{P} for example), the excess pressure supplied
by the field has simply increased the oscillation frequency of the density
contrast.

Conclusively, the presence of a cosmological magnetic field during the radiation
era does not cause significant changes in the evolutionary patterns of the
density gradients. However, as far as their actual growth is concerned, the
field impact is evidently negative, although still secondary to the effects
induced by the pressure of the relativistic matter, which still dominates
their evolution.

\subsection{The Dust Era}
%%%%%%%%%%%%%%%%%%%%%%%%%
After the radiation era ends, dust dominates and the energy density evolves as
$\mu=\Bbb{M}_D/S^3$ with $\dot{\Bbb{M}}_D=0$. Thus, for vanishing cosmological
constant the scale factor changes as $S=\alpha t^{2/3}$, where $t$ measures the
observer's time and $\alpha\equiv(3\kappa\Bbb{M}_Dc^4/4)^{1/3}$. Also,
$\Theta=2/t$ and $\mu=4/3\kappa c^4t^2$. During this period the Alfv\'{e}n
velocity falls as $c_A^2=\Bbb{E}/\alpha t^{2/3}$, with $\dot{\Bbb{E}}=0$,
reflecting the fact that the magnetic energy density drops faster than that of
the matter. So, relative to a reference frame comoving with the expanding
fluid, equations (\ref{nlbddDel})-(\ref{nlbdcB}) become
\begin{equation}
\frac{d^2\Delta^{(\nu)}}{dt^2}=-\frac{4}{3t}\frac{d\Delta^{(\nu)}}{dt}+
\frac{2}{3t^2}\Delta^{(\nu)}+
\frac{8\Bbb{E}}{9c^2\alpha t^{8/3}}{\cal K}^{(\nu)}-
\frac{2\Bbb{E}}{3c^2\alpha t^{8/3}}
\left(1+\frac{3\nu^2c^2 t^{2/3}}{4\alpha^2}\right){\cal B}^{(\nu)} ,
\label{dnlbddDel}
\end{equation}
\begin{equation} \frac{d{\cal
K}^{(\nu)}}{dt}=\frac{4\Bbb{E}}{3c^2\alpha t^{5/3}}{\cal B}^{(\nu)} ,
\label{dnlbdcK} \end{equation}
\begin{equation}
\frac{d{\cal B}^{(\nu)}}{dt}=\frac{4}{3}\frac{d\Delta^{(\nu)}}{dt} .
\label{dlbdcB}
\end{equation}

On superhorizon scales (i.e. $\nu\rightarrow0$, or equivalently
$\nu^2c^2t^{2/3}/\alpha^2\ll1$)\footnote{The post-equilibrium evolution of the
scale factor implies that the long-wavelength condition
$\lambda_{phys}\gg d_H$ translates into $\nu^2c^2t^{2/3}/\alpha^2\ll1$.} the
above system simplifies into
\begin{equation}
\frac{d^2\Delta^{(\nu)}}{dt^2}=-\frac{4}{3t}\frac{d\Delta^{(\nu)}}{dt}+
\frac{2}{3t^2}\Delta^{(\nu)}+
\frac{8\Bbb{E}}{9c^2\alpha t^{8/3}}{\cal K}^{(\nu)}-
\frac{2\Bbb{E}}{3c^2\alpha t^{8/3}}{\cal B}^{(\nu)} ,  \label{LdnlbddDel}
\end{equation}
\begin{equation}
\frac{d{\cal K}^{(\nu)}}{dt}=
\frac{4\Bbb{E}}{3c^2\alpha t^{5/3}}{\cal B}^{(\nu)} ,
\label{LdnlbdcK}
\end{equation}
\begin{equation}
\frac{d{\cal B}^{(\nu)}}{dt}=\frac{4}{3}\frac{d\Delta^{(\nu)}}{dt} .
\label{LdnlbdcB}
\end{equation}
To obtain an analytic solution, we assume that the perturbed universe has flat
spatial sections (i.e. ${\cal K}=0$) and also consider the special case where
${\cal B}^{(\nu)}=4\Delta^{(\nu)}/3$. Then, we are left with the following
differential equation\footnote{According to equation (\ref{LdnlbdcB}), the
condition ${\cal B}^{(\nu)}/\Delta^{(\nu)}=4/3$ requires that the same ratio
holds at the initial moment too. In other words, solutions (\ref{LdDel}) and
(\ref{LdDel1}) presume that, as the large-scale spatial variations in the
magnetic and the fluid energy densities enter the post-equilibrium era, their
ratio equals $4/3$. Such a simplifying step is not unreasonable at all since,
as equation (\ref{rnlbdcB}) suggests, $B^{(\nu)}\sim\Delta^{(\nu)}$ by the end
of the radiation era.}
\begin{equation}
\frac{d^2\Delta^{(\nu)}}{dt^2}=-\frac{4}{3t}\frac{d\Delta^{(\nu)}}{dt}+
\frac{2}{3t^2}\left(1-\frac{4\Bbb{E}}{3c^2\alpha t^{2/3}}\right)\Delta^{(\nu)} .
\label{LdnlbddDel1}
\end{equation}
Notice that at later times (i.e. $t\rightarrow\infty$) the magnetic term in the
parenthesis becomes completely irrelevant. So, in agreement with \cite{TB}, we
recover the power-law evolution
\begin{equation}
\Delta^{(\nu)}=\Delta^{(\nu)}_{-}t^{-1}+\Delta_{+}^{(\nu)}t^{2/3} ,
\label{LdDel}
\end{equation}
also familiar from the study of a non-magnetized cosmological model. The
alternative early time solution
\begin{eqnarray}
\Delta^{(\nu)}(t)&=&
\left[\Delta_1^{(\nu)}\sin\left(\frac{\epsilon}{t^{1/3}}\right)+
\Delta_2^{(\nu)}\cos\left(\frac{\epsilon}{t^{1/3}}\right)\right]
\epsilon t^{1/3}+
\nonumber\\\nonumber\\&\mbox{}&
\left[\Delta_1^{(\nu)}\cos\left(\frac{\epsilon}{t^{1/3}}\right)-
\Delta_2^{(\nu)}\sin\left(\frac{\epsilon}{t^{1/3}}\right)\right]
\left(t^{2/3}-\frac{\epsilon^2}{3}\right),
\label{LdDel1}
\end{eqnarray}
where $\epsilon\equiv 2\sqrt{2\Bbb{E}}/c\sqrt{\alpha}$, suggests that under the
magnetic influence the long-wavelength aggregations of the material component
oscillate with an amplitude that increases as $t^{2/3}$.

On scales well below the horizon (i.e. $\nu\rightarrow\infty$, or equivalently
$\nu^2c^2t^{2/3}/\alpha^2\gg1$), equations (\ref{dnlbddDel})-(\ref{dlbdcB})
become
\begin{equation}
\frac{d^2\Delta^{(\nu)}}{dt^2}=-\frac{4}{3t}\frac{d\Delta^{(\nu)}}{dt}+
\frac{2}{3t^2}\Delta^{(\nu)}+
\frac{8\Bbb{E}}{9c^2\alpha t^{8/3}}{\cal K}^{(\nu)}-
\frac{\nu^2\Bbb{E}}{2\alpha^3t^2}{\cal B}^{(\nu)} ,
\label{SdnlbddDel}
\end{equation}
\begin{equation}
\frac{d{\cal K}^{(\nu)}}{dt}=\frac{4\Bbb{E}}{3c^2\alpha t^{5/3}}
{\cal B}^{(\nu)} ,  \label{SdnlbdcK}
\end{equation}
\begin{equation}
\frac{d{\cal B}^{(\nu)}}{dt}=\frac{4}{3}\frac{d\Delta^{(\nu)}}{dt} .
\label{SdnlbdcB}
\end{equation}
Again, by ignoring any effects from the spatial curvature we obtain the
following power-law evolution for the density contrast
\begin{equation}
\Delta^{(\nu)}(t)=\Delta_1^{(\nu)}t^{z_1}+\Delta_2^{(\nu)}t^{z_2} ,
\label{SdDel}
\end{equation}
with,
\begin{equation}
z_{1,2}=-\frac{1}{6}
\left(1\pm5\sqrt{1-\frac{24\nu^2\Bbb{E}}{25\alpha^3}}\right) .  \label{dz}
\end{equation}
Notice that in the absence of the magnetic field (i.e. when $\Bbb{E}=0$) we are
left with the well known solution (i.e. $z_{1,2}=2/3$, $-1$) of the
magnetic-free case.  In quantitative agreement with Ruzmaikina and Ruzmaikin
we find that the field presence reduces the growth rate of the inhomogeneities
proportionally to the ratio $t^{2/3}H^2/\mu c^2$.

We conclude by arguing that the presence of a cosmological magnetic field always
opposes the growth of matter aggregations, either by forcing them to oscillate
or by reducing their growth rate. The magnetic influence ceases only at the
later stages of the dust era, when the relative strength of the field becomes
negligibly small. It should be emphasized that result (\ref{SdDel}) refers
to wavelengths that lie within the horizon but are much larger than the
Jeans length at the time. Otherwise the pressure effects of the ordinary
non-relativistic matter become important preventing the density gradients from
growing. This fact, together with the oscillatory nature of solution
(\ref{LdDel1}), suggests that earlier in the dust era any actual growth is
confined to scales comparable to the horizon size at the time. 

\section{Conclusions}
%%%%%%%%%%%%%%%%%%%%%
We have explored the influence of a primordial magnetic field upon the
kinematical and the dynamical evolution of perturbed cosmological models
containing perfect fluids with non-vanishing pressure. We employed the
Ellis-Bruni covariant and gauge-invariant formalism, first applied to the
analysis of magnetized cosmologies in \cite{TB}, to derive the full set of
equations determining the linear evolution of an almost-FFRW universe
containing a perfectly conducting medium. Relative to the dust era examined in
\cite{TB}, the principal new complexities are due to changes in the observer's
motion under the simultaneous action of the perturbed medium and the magnetic
field. These changes are best seen in the different form of the momentum density
conservation law (see eqn (\ref{lmdc})), which in turn implies a modified
acceleration for the fluid. In fact, this is the reason for essentially all
the extra complications in the evolutionary patterns of the pre-equilibrium
era. We have quantified the magnetohydrodynamical effects upon the kinematics
and the dynamics of a universe dominated by a barotropic perfect fluid. We
found an acceleration that depends on density gradients as well as on the
gradients of the field. It is no longer normal to the field vector and can have
subtle effects upon the evolution of fundamental cosmological parameters. Of
particular interest is the first order magneto-geometrical contribution to the
deceleration parameter. We show that, unlike ordinary matter which always slows
the expansion down, the magnetic field can act as a driving force through its
interaction with the geometry of the spatial sections. An analogous effect is
found upon the Ricci scalar of the observer's instantaneous rest space. On
large scales, the magnetic field tends to smooth out the curvature of the
spacelike surfaces and restore their initial flatness.

As in \cite{TB}, we were primarily interested in studying the growth of density
inhomogeneities in a magnetized environment. Here, we have defined four scalar
variables that measure spatial variations in the energy densities of the medium
($\Delta$) and the magnetic filed (${\cal B}$), spatial inhomogeneities in the
expansion (${\cal Z}$), and deviations from the spatial flatness of the
background universe (${\cal K}$). We provide a system of four linear
first order differential equations that describes the evolution of these
disturbances and ultimately dictates the behaviour of spatial
matter aggregations. We have obtained analytic solutions both at the long and
at the short-wavelength limit during the radiation and the dust eras. In
\cite{TB} we argued for the relative unimportance of the field during the
dust era and on scales that exceed the horizon at the time. Here,
we were able to confirm and also refine those results. More specifically,
we have found that any magnetic effects upon long-wavelength
matter aggregations cease completely as the dust era enters its later
stages. During this period the inhomogeneities grow exactly as those in a
non-magnetized universe. Soon after equilibrium however, the extra pressure
of the field could have forced the density gradients to oscillate, thus
preventing them from growing. Nevertheless, the weakness of the field means
that such large-scale oscillations are short-lived and that the epoch of
unimpeded growth begins almost immediately after equilibrium. On scales smaller
than the horizon, but larger than the associated at the time Jeans length, the
disturbances undergo a power-law growth but at a slower pace relative to the
magnetic-free case. The field pressure also affects the evolution of the
density contrast during the radiation era. Here, it adds to the pressure of the
relativistic matter and impedes any further gravitational clumping of the
medium. On large scales we have found that the field inhibits the growth of
the inhomogeneities by an amount proportional to its relative strength, whereas
on subhorizon regions it increases the frequency of their oscillations. In the
radiation era the magnetic effect supplements that from the pressure of the
relativistic matter. During this period, the fate of small-scale inhomogeneities
is affected by plasma processes \cite{SB1}.

\section*{Acknowledgments}
%%%%%%%%%%%%%%%%%%%%%%%%%%
C.G. Tsagas was supported by the Greek State Scholarship Foundation and J.D
Barrow by a PPARC Senior Fellowship.  The authors also wish to thank Roy
Maartens, Marco Bruni and Kandu Subramanian for helpful discussions.

\section*{Appendices}
%%%%%%%%%%%%%%%%%%%%%
\appendix

\section{Gauge-invariance for the 3-gradients of Spatial Vectors} 
%%%%%%%%%%%%%%%%%%%%%%%%%%%%%%%%%%%%%%%%%%%%%%%%%%%%%%%%%%%%%%%%%
In \cite{TB} (see Appendix A there) it was stated that the metric of an exact
FRW or Bianchi-I spacetime can always be brought in the diagonal form
\begin{equation}
g_{ij}=\mbox{\rm diag}(g_{00}, g_{11}, g_{22}, g_{33}) ,  \label{gij}
\end{equation}
with its components being functions of proper-time only (i.e. $g_{ii}=g_{ii}(t)$
- no summation over $i$). Based on this we then proceeded to prove the
gauge-invariance of ${\cal M}_{ij}$, the spatial tensor that describes the
variations of the magnetic field vector as seen by two neighbouring fundamental
observers. However, though our initial statement is correct within a Bianchi-I
and a spatially flat FRW cosmology (see for example \cite{S} for verification),
it can not be extended to spatially curved FRW spacetimes. So, the
gauge-independence of the magnetic 3-gradients has been established simply
within perturbed FFRW and Bianchi-I universes.  What we actually showed in
\cite{TB} was that the spatial flatness of the aforementioned spacetimes
(together with their zero rotation) is sufficient for the 3-gradients of any
homogeneous spacelike vector field to be gauge-invariant. Next we argue that
within the limits of a FRW model such a requirement is also necessary. Indeed,
in the observer's rest space the Ricci identity takes the form (see Appendix B)
\begin{equation}
\mbox{}^{(3)}\nabla_{[i}\mbox{}^{(3)}\nabla_{j]}v_k=
-\frac{1}{c^2}\omega_{ij}h_k^{\hspace{1mm}q}\dot{v}_q+
\frac{1}{2}\mbox{}^{(3)}R_{qkji}v^q ,  \label{vcom}
\end{equation}
where $v_iu^i=0$.  The above equation, which is presented here in its exact
form, clearly states that in a non-rotating spacetime (i.e. $\omega_{ij}=0$) the
3-gradients of any spacelike vector vanish (i.e. $\mbox{}^{(3)}\nabla_iv_j=0$)
only when $\mbox{}^{(3)}R_{ijkq}v^i=0$. Contracting the latter over the indices
$j$ and $q$ we obtain the new restriction $\mbox{}^{(3)}R_{ij}v^j=0$. In a FRW
spacetime $\mbox{}^{(3)}R_{ij}=\mbox{}^{(3)}Rh_{ij}/3$, which means that the
gauge-invariance of $\mbox{}^{(3)}\nabla_iv_j$ requires $\mbox{}^{(3)}R$ to
vanish. This in turn ensures that $\mbox{}^{(3)}R_{ijkq}=0$ and therefore the
spatial flatness of the model. Clearly, the introduction of 3-vectors into the
spatially isotropic Friedmannian cosmologies is only an
approximation. Nevertheless, the 3-gradients of such a homogeneous spacelike
vector, cannot be treated as gauge-independent variables unless the FRW
cosmology is spatially flat.

%When dealing with a general non-rotating spacetime, equation (\ref{vcom})
%implies that the spatial gradients of any 3-vector $v_i$ are not gauge-invariant
%if $\mbox{}^{(3)}R_{ij}v^j\neq0$. On the other hand, a spacetime metric such as
%(\ref{gij}), together with zero vorticity, is sufficient for the spatial
%gradients of any homogeneous spacelike vector to vanish. Combining these two
%statements we may argue that: the metric of a non-rotating spacetime $W$
%cannot be brought into a diagonal form with all its components being functions
%of proper-time only, if $\mbox{}^{(3)}R_{ij}v^j\neq0$ for some homogeneous
%spacelike vector $v_i$.

\section{Auxiliary Relations}
%%%%%%%%%%%%%%%%%%%%%%%%%%%%%
Following \cite{EBH} we point out that generally the operator
$\mbox{}^{(3)}\nabla_i$ cannot be treated as the standard covariant derivative
of a 3-dimensional hypersurface because in a rotating spacetime the {\it
defect tensor} does not vanish. Thus one cannot assume the usual commutation
relations but should use expressions that include possible rotational terms.
A selection of such formulae can be found in \cite{EBH} (see Appendix A
there). Here we present only those essential to our analysis.

Commutations between the spatial gradients of scalars and spacelike vectors
are respectively given by 
\begin{equation}
\mbox{}^{(3)}\nabla_{[i}\mbox{}^{(3)}\nabla_{j]}f=
-\frac{1}{c^2}\omega_{ij}\dot{f} ,  \label{scmtr}
\end{equation}
and
\begin{equation}
\mbox{}^{(3)}\nabla_{[i}\mbox{}^{(3)}\nabla_{j]}v_k=
-\frac{1}{c^2}\omega_{ij}h_k^{\hspace{1mm}q}\dot{v}_q+
\frac{1}{2}\mbox{}^{(3)}R_{qkji}v^q ,  \label{vcmtr}
\end{equation}
where $f$ can be any scalar and $v_i$ is a spatial vector (i.e.
$v_iu^i=0$). Equation (\ref{vcmtr}) is also regarded as the general expression
of the 3-Ricci identity. Commutations between the spatial gradients and the time
derivatives of these quantities are governed by
\begin{equation}
\mbox{}^{(3)}\nabla_i\dot{f}-
h_i^{\hspace{1mm}j}\left(\mbox{}^{(3)}\nabla_jf\right)^{\cdot}=
-\frac{1}{c^2}\dot{f}a_i+\frac{1}{3}\Theta\mbox{}^{(3)}\nabla_if+
\mbox{}^{(3)}\nabla_jf
\left(\sigma^j_{\hspace{1mm}i}+\omega^j_{\hspace{1mm}i}\right) ,  \label{scmtr1}
\end{equation}
and
\begin{equation}
\mbox{}^{(3)}\nabla_i\dot{v}_j-h_i^{\hspace{1mm}k}h_j^{\hspace{1mm}q}
\left(\mbox{}^{(3)}\nabla_kv_q\right)^{\cdot}=
\frac{1}{3}\Theta\mbox{}^{(3)}\nabla_iv_j ,  \label{vcmtr1}
\end{equation}
where the latter appears here in its linearized form and applies only to
first-order (i.e. $v_i\equiv0$ in the background) spacelike vectors. Commutator
(\ref{vcmtr1}) provides an additional first order relation, which plays an
important role in our analysis. In particular, assuming that
$\mbox{}^{(3)}\nabla_if$ vanishes in the background, we may linearize the
3-divergence 0f (\ref{scmtr1})) to obtain
\begin{equation}
\left(\mbox{}^{(3)}\nabla^2f\right)^{.}-\mbox{}^{(3)}\nabla^2\dot{f}=
\frac{1}{c^2}\dot{f}A-\frac{2\Theta}{3}\mbox{}^{(3)}\nabla^2f ,  \label{scmtr2}
\end{equation}
recalling that $A=\mbox{}^{(3)}\nabla_ia^i$ to first order. The above is used
to derive the evolution formula of ${\cal B}$, the scalar that describes
spherically symmetric changes in the energy density of the magnetic field.


\begin{thebibliography}{10}
%%%%%%%%%%%%%%%%%%%%%%%%%%%
\bibitem[1]{TB} Tsagas C.G. and Barrow J.D. (1997), {\it Classical and
Quantum Gravity}, {\bf 14}, 2539.

\bibitem[2]{EB} Ellis G.F.R. and Bruni M. (1989), {\it Phys. Rev. D},
{\bf 40}, 1804.

\bibitem[3]{RR} Ruzmaikina T.V. and Ruzmaikin A.A (1971), {\it Soviet Astr.},
{\bf 14}, 963.

\bibitem[4]{W} Wasserman I. (1978), {\it Ap. J.}, {\bf 224}, 337.

\bibitem[5]{SB1} Subramanian K. and Barrow J.D. (1997), astro-ph/9712083.

\bibitem[6]{SB2} Subramanian K. and Barrow J.D. (1998), astro-phy/9803061.

\bibitem[7]{BFS} Barrow J.D., Ferreira P.G. and Silk J. (1997),
{\it Phys. Rev. Lett.}, {\bf 78}, 3610.

\bibitem[8]{B} Barrow J.D. (1997), {\it Phys. Rev. D}, {\bf 55}, 7451.

\bibitem[9]{Eh} Ehlers J. (1961), {\it Abh. Mainz Akad. Wiss. Lit.},
{\bf 11}, 1.

\bibitem[10]{E2} Ellis G.F.R. (1973), in {\it Carg\`ese Lectures in Physics,
Vol. I}, ed. E. Schatzmann, Gordon and Breach, New York, 1.

\bibitem[11]{M} Maartens R. (1997), {\it Phys. Rev. D}, {\bf 55}, 463.

\bibitem[12]{MB} Maartens R. and Bassett B.A.C.C. (1998),
{\it Class. Quantum Grav.}, {\bf 15}, 705.

\bibitem[13]{SW} Stewart J.M. Walker M. (1974), {\it Proc. Roy. Soc. London A},
{\bf 341}, 49.

\bibitem[14]{EBH} Ellis G.F.R., Bruni M. and Hwang J. (1990),
{\it Phys. Rev. D}, {\bf 42}, 1035.

\bibitem[15]{T} C.G. Tsagas (1998), {\it PhD Thesis}, University of Sussex.

\bibitem[16]{H} Hawking S.W. (1966), {\it Ap. J.}, {\bf 145}, 544.

\bibitem[17]{EHB} Ellis G.F.R., Hwang J. and Bruni M. (1989),
{\it Phys. Rev. D}, {\bf 40}, 1819.

\bibitem[18]{BDE} Bruni M., Dunsby P.K.S. and Ellis G.F.R. (1992),
{\it Ap. J.}, {\bf 395}, 34.

\bibitem[19]{DBE} Dunsby P.K.S., Bruni M. and Ellis G.F.R. (1992),
{\it Ap. J.}, {\bf 395}, 54.

\bibitem[20]{Ha} Harrison E.R. (1967), {\it Rev. Mod. Phys.}, {\bf 39}, 862.

\bibitem[21]{AS} Abramowitz M. and Stegun I.A. (1984), {\it Pocketbook of
Mathematical Functions}, Verlag Harri Deutsch, Thun-Frankfurt am Main.

\bibitem[22]{Tu} Turnbull H.W. (1952), {\it Theory of Equations}, Oliver and
Boyd, Edinburgh.

\bibitem[23]{P} Padmanabhan T. (1993), {\it Structure Formation in the
Universe}, Cambridge University Press, Cambridge.

\bibitem[24]{S} Stephani H. (1990), {\it General Relativity}, Cambridge
University Press, Cambridge.

\end{thebibliography}
\end{document}